\documentclass{WileyMSP-template}

\begin{document}

\pagestyle{fancy}
\rhead{\includegraphics[width=2.5cm]{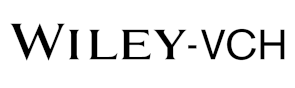}}

\title{Resistively-detected NMR lineshapes in a local filling $\nu < 1$ quantum Hall breakdown}

\maketitle

\author{M.~H.~Fauzi}

\author{T. Sobue}

\author{A. Noorhidayati}


\author{K. Sato}

\author{K. Hashimoto}

\author{Y. Hirayama}



\begin{affiliations}
M.~H.~Fauzi\\
Research Center for Physics, National Research and Innovation Agency, South Tangerang, Banten 15314, Indonesia

T. Sobue, A. Noorhidayati, K. Sato, K. Hashimoto\\
Department of Physics, Tohoku University, Sendai 980-8578, Japan\\

Y. Hirayama\\
Center for Spintronics Research Network, Tohoku University, Sendai 980-8577, Japan\\
Center for Science and Innovation in Spintronics (Core Research Cluster), Tohoku University, Sendai 980-8577, Japan

\end{affiliations}



\begin{abstract}

Resistively-detected NMR (RDNMR) is a unique characterization method enabling highly-sensitive NMR detection for a single quantum nanostructure, such as a quantum point contact (QPC). In many studies, we use dynamic nuclear polarization and RDNMR detection in a quantum Hall breakdown regime of a local QPC filling factor of 1 ($\nu_{\rm{qpc}} = 1$). However, the lineshapes of RDNMR signal are proved to be complicated and still not fully understood yet. Here, we systematically polarize the nuclear spins by current-pumping from the close vicinity of $\nu_{\rm{qpc}} = 1$ conductance plateau all the way down to pinch-off point, providing a clear evidence that the spin-flip scattering between two edge channels at the lowest Landau level still occurs in the constriction even when it is close to the pinch off point ($G \approx 10^{-2}$ $2e^{2}/h$). The collected RDNMR spectra reveal two sets of distinguished features. First, in a strong to intermediate tunneling regime, we observe an ordinary resistance dip lineshape but interestingly its transition frequency follows a snake-like pattern, an indicative of spatial modulation of electron density in the QPC. Second, in a weak tunneling regime, the spectrum turns into a dispersive lineshape which we interpret due to the build up of two sets of nuclear spin polarization that are in contact with different electron spin polarization.

\end{abstract}


\section{Introduction}

Halperin, in the early 80s, has predicted the importance of edge states to the transport properties in the quantum Hall effect\cite{Halperin1982}. However, it is not until the late 80s that the importance of edge states has gained traction and accepted widely by the community \cite{Haug1993}. Since then, the chiral nature of edge states has been exhaustedly studied due to its potential application for future electronics \cite{Duprez2019}. In the early 90s, it has been recognized that the nuclear spin degree of freedom resides in GaAs crystal can influence quantum Hall edge transport \cite{Kane1992, Wald1994, Dixon1997}. The nuclear spin and electron spin in GaAs are glued by the so-called hyperfine interaction \cite{Hirayama2009}. A non-zero nuclear spin polarization would exert an extra magnetic field for electrons, known as Overhauser field. Likewise, a net electron spin polarization would create an effective magnetic field (Knight field) and shift the nuclear energy level.

Since the pioneering work of Wald \textit{et al.} in early 90s \cite{Wald1994}, resistively-detected NMR (RDNMR) has emerged as a powerful tool to study characteristics of a quasi-1D channel defined by a quantum point contact (QPC). The characteristics, which cannot be detected by any conventional transport measurements, such as ultra-low strain distribution\cite{Fauzi2019}, have been successfully detected as discussed in the next section. In these studies, we set the bulk 2DEG outside of the QPC at the filling of $\nu_{\rm{b}}= 2$ and inside filling factor in the QPC to be less than 1 ($\nu_{\rm{qpc}} < 1$). The electron tunneling between edge channels with differing spin polarity near the QPC determines dynamic nuclear polarization and RDNMR signal\cite{Wald1994, Dixon1997, Fauzi2017}. Although the RDNMR signal is mainly dominated by positive nuclear polarization, i.e. parallel to the applied magnetic field, due to forward edge-channel scattering events when $\nu_{\rm{qpc}}$ is tuned slightly lower than 1 \cite{Fauzi2019, Fauzi2017, Fauzi2018, Annisa2020}. However, the negative nuclear polarization also plays a role and a complicated dispersive line-shape signal appears in some conditions\cite{Fauzi2017}. The observed dispersive lineshapes have been reported in a 2D system \cite{Desrat2002, Tracy2006, Kodera2006, Dean2009, Bowers2010, Desrat2013, Desrat2015, Piot2016} and are interpreted due to the formation of spin unpolarized puddles in the sea of spin-polarized background \cite{Desrat2013}.

Previous RDNMR studies are overwhelmingly reported in a single quantum point contact device and in a strong tunneling regime between the two edge channels at the lowest Landau level \cite{Keane2011, Chida2012, Fauzi2017, Okazaki2018, Hashisaka2020}. In attempt to comprehensively study local hyperfine-mediated transport in the lowest Landau level, in this study, we investigate RDNMR spectra in a wide range of local filling factor $0 < \nu_{\rm{qpc}} < 1$, encompassing strong to weak tunneling regime between the inner and outer edge channel at the lowest Landau level. Furthermore, we use a QPC device with multiple gate structures to control a potential shape in and near the QPC channel. We will discuss these newly obtained results using this device after discussing basic characteristics of RDNMR in a QPC device.




\section{Basic of RDNMR in a quantum point contact}

\begin{figure*}[t]
\begin{center}    
\centering
\includegraphics[width=
\linewidth]{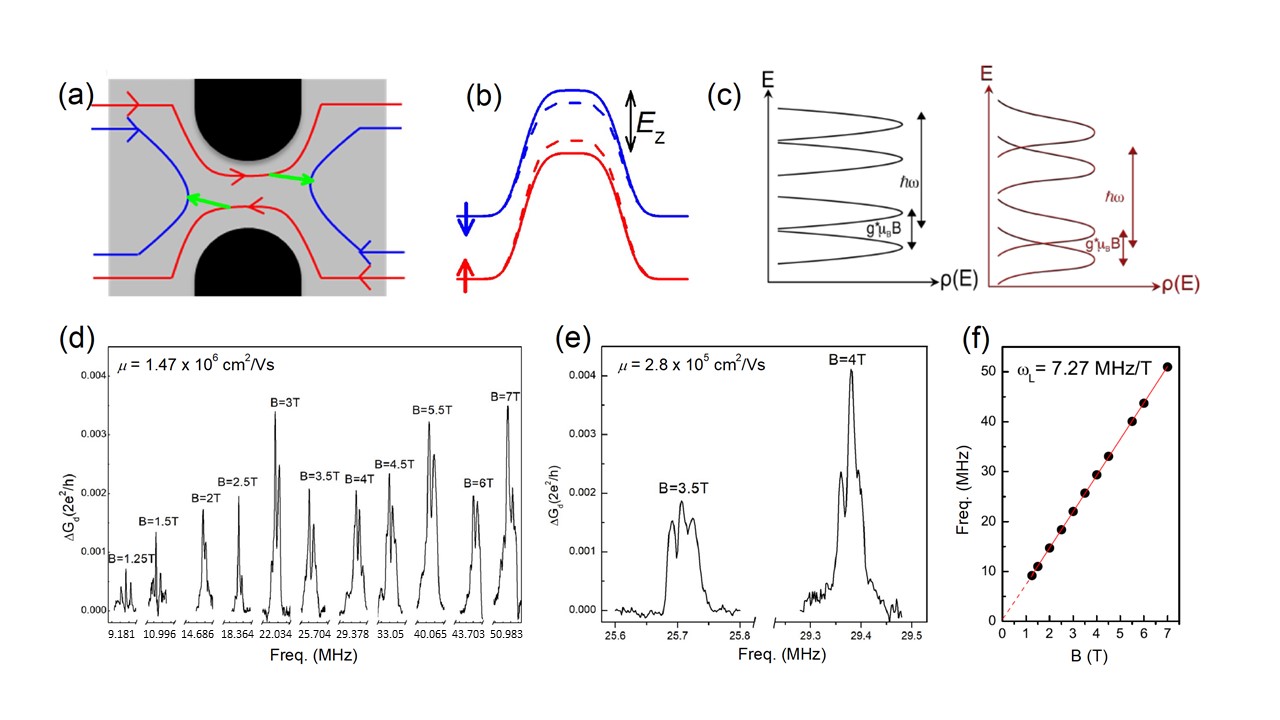}
\end{center}
\caption{(a) Schematic of forward spin flip scattering indicated by the green arrow between up (red line) and down-spin (blue line) edge channel in a quantum point contact. (b) The potential barrier seen by electrons is modified in the presence of a positive nuclear spin polarization (dashed lines). The effective electronic Zeeman energy felt by electrons is reduced. (c) Schematic of Landau levels with broadening when the spin is fully and partially resolved. (d) $^{75}$As RDNMR spectra measured in high-mobility ($\mu = 1.47 \times 10^{6}$ cm$^2$/Vs). The horizontal scale denotes the center of resonance line. (e). $^{75}$As RDNMR spectra measured in low-mobility devices ($\mu = 2.8 \times 10^{5}$ cm$^2$/Vs). (f) $^{75}$As resonance frequency plotted as a function of magnetic field. The measured slope is $7.27$ MHz/T. (reproduced with permission from ref. \cite{Annisa2020})}
\label{Fig01} 
\end{figure*}

The simplest way to dynamically generate and detect an ensemble of nuclear spin polarization in a QPC is by setting the bulk filling factor to $\nu_{\rm{b}} = 2$ and the filling of the point contact to $\nu_{\rm{qpc}} < 1$. This technique goes all the way back to the pioneering work of Wald \textit{et al.} in early 90s \cite{Wald1994}. When a sufficient bias current flows through, forward spin flip scattering events marked by the green arrow in Fig. \ref{Fig01}(a) is enhanced. A hyperfine interaction, particularly strong in GaAs semiconductor \cite{Hirayama2009}, allows an effective exchange of angular momentum between an electron and a nuclear spin. The momentum exchange satisfies the energy conservation. The scattering events can mediate dynamic nuclear polarization (DNP) in the QPC. The polarized nuclei built up in the QPC modifies the electronic Zeeman energy and hence influences the electron transport through the saddle point potential. For a positive nuclear polarization generated by the forward spin flip scattering, the effective Zeeman energy will be reduced as schematically displayed Fig. \ref{Fig01}(b) in the case of GaAs. It means in the presence of the positive nuclear spin polarization, the spin-up electrons would see an increase in the barrier potential. During the build up of nuclear spin polarization by current-induced DNP, the resistance will increase with time typically over a few hundred seconds \cite{Fauzi2017} and it will reach a steady-state condition where the rate of spin flip-flop and the loss due to nuclear spin relaxation and diffusion get balanced \cite{Singha2017}.

Knowing how the DNP works and how it influence the transport, it is then relatively easy to understand how the RDNMR spectrum would look like. In a continuous wave experiment, where the in-plane radio frequency is swept at a slow speed so that the nuclear and electron spin are in equilibrium, the resistance would drop when it hits the Larmor frequency of a nuclei and go back up after it passes the resonance point where the nuclear spins are getting repolarized.

The DNP and RDNMR in the situation mentioned above depends on the forward spin flip scattering events between the spin-resolved edge channels marked by the green arrow in Fig. \ref{Fig01}(a). Naturally, a well-defined spin-resolved edge channel becomes an important factor for achieving an effective DNP process and successful RDNMR detection. The electron mobility of the GaAs quantum well, on which the QPCs are fabricated, determines a broadening of the spin-resolved Landau-level as shown in Fig. \ref{Fig01}(c) so that the minimum magnetic field necessary for the RDNMR detection increases with decreasing the mobility \cite{Annisa2020}. Typical RDNMR signals obtained for high-mobility ($\mu = 1.47 \times 10^{6}$ cm$^2$/Vs at electron density of $1.8 \times 10^{11}$ cm$^{-2}$) and low-mobility ($\mu =2.8 \times 10^{5}$ cm$^2$/Vs at electron density of $1.8 \times 10^{11}$ cm$^{-2}$) QPCs are shown in Fig. \ref{Fig01}(d) and (e), respectively. They are obtained by setting $\nu_{\rm{b}} = 2$ and $\nu_{\rm{qpc}}$ slightly less than 1, corresponding to a strong tunneling regime discussed later. For the high-mobility QPCs, we could clearly observe the RDNMR signal down to $B = 1.25$ T. For the low-mobility QPCs, successful observation of the signal is limited to $B > 3$ T. However, all signals reflect the Larmor frequency of $^{75}$As as shown in Fig. \ref{Fig01}(f) and, more importantly, both lineshape and broadening showed the same tendency, independent of the mobility \cite{Annisa2020}. It means that we can see intrinsic characteristics of the RDNMR signals even for the low-mobility QPC device when we set the magnetic field high enough so that the up and down spin edge channels are well resolved.

\begin{figure}[t]
\begin{center}    
\centering
\includegraphics[width=0.7
\linewidth]{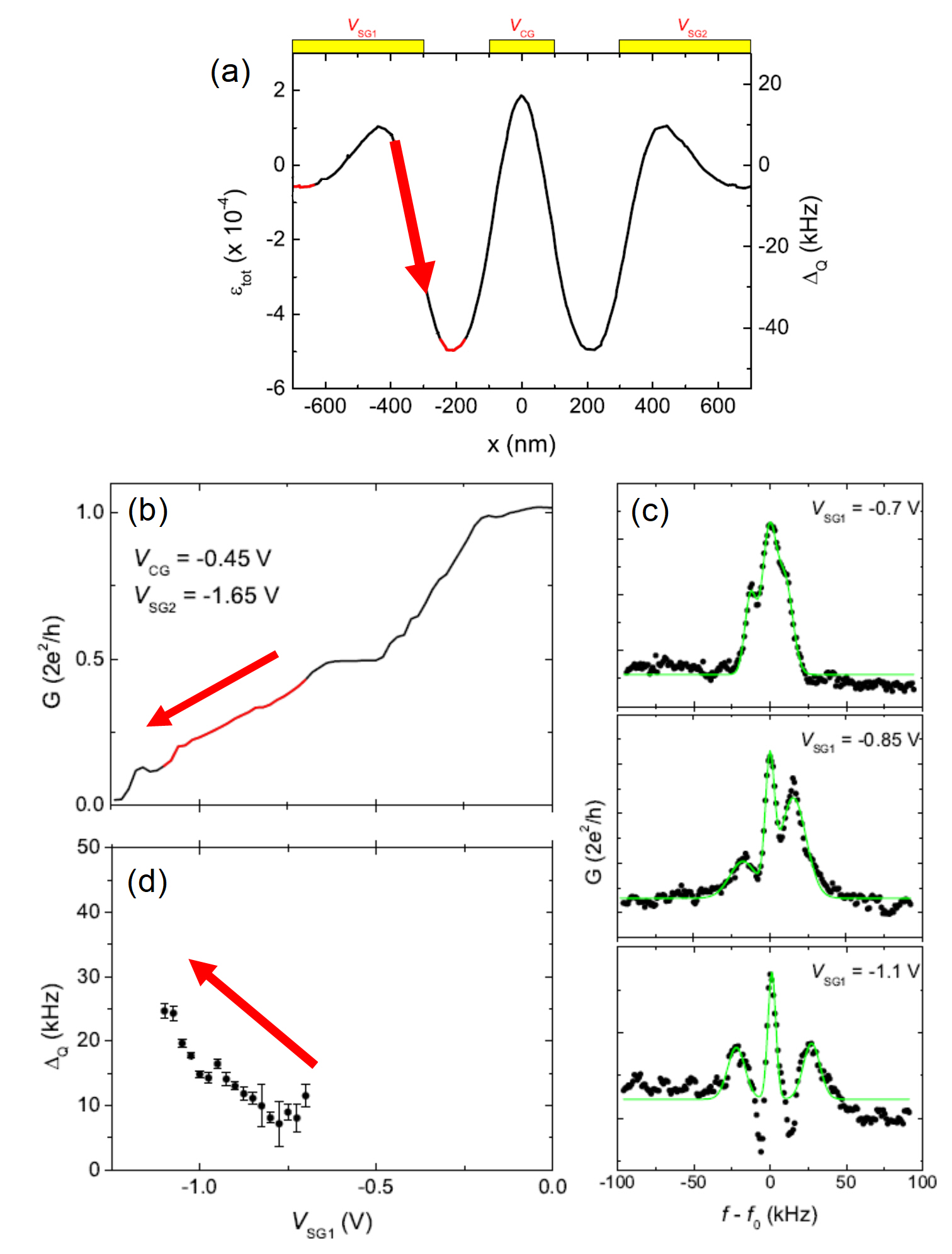}
\end{center}
\caption{(a) Calculated total strain field $\epsilon_{\rm{tot}}$ and the corresponding quadrupole splitting $\Delta_{\rm{Q}}$ felt by $^{75}$As nuclei located 175 nm below the surface. The region of interest is marked by the red arrow. (b) Diagonal conductance trace as a function of the left-hand-side split gate $V_{\rm{SG1}}$. The right-hand-side gate $V_{\rm{SG2}}$ and center gate $V_{\rm{CG}}$ are set to $-1.65$ V and $-0.45$ V in order to detect the strain field indicated by the red arrow in panel (a). (c) $^{75}$As RDNMR spectra selected along the red conductance trace in panel (b). (d) Quadrupole splitting from the RDNMR spectra collected along the red line in panel (b). (reproduced with permission from ref. \cite{Fauzi2019})}
\label{Fig02}
\end{figure}

Another important feature observed in the RDNMR signals is the quadrupolar splitting. The $^{75}$As has $I = 3/2$ and the nuclear spin levels are energetically separated into four states, resulting in three RDNMR peaks if the quadrupolar splitting is large enough. The quadrupolar splitting reflects strain field felt by the nuclear spin and its high-sensitivity enables us to detect small strain difference on the order of $\epsilon_{\rm{tot}}$ of $10^{-4}$ [8]. Figure \ref{Fig02}(a) shows calculated total strain $\epsilon_{\rm{tot}}$ distribution in the QPC device. The strain becomes maximum in between the metal gates and decreases under the gate. In the RDNMR, the DNP occurs by the flip-flop process between spins of itinerant electrons and local nuclear spins. Moreover, the RDNMR signal is determined by the transport characteristics through a saddle potential of a QPC. Therefore, the quadrupolar splitting of the RDNMR signals reflects the strain in the electron channel of the QPCs. In the case of the experiments shown in Fig. 2 \cite{Fauzi2019}, when we applied more a negative voltage (Vsg1) to the split gate 1 (SG1), the electron channel moved from beneath the SG1 to the region in between the gates as shown by the red arrow in Fig. \ref{Fig02}(a). Correspondingly, the QPC conductance decreased as shown in Fig. \ref{Fig02}(b). The typical RDNMR signals are shown in Fig. \ref{Fig02}(c) and the quadrupolar splitting in Fig. \ref{Fig02}(d) indicates a clear enhancement corresponding to an increase in the strain felt by the nuclei. This means that we can infer a position of the channel from the measured quadruppolar splitting.

Finally, we should discuss the Knight shift coming from a net electron spin polarization. In the $\nu_{\rm{qpc}} \sim 1$ regime, itinerant electrons in the QPC are expected to be spin polarized and we can expect a slight shift of RDNMR resonant frequency due to the Knight shift. The Knight shift is proportional to the number of spin-polarized electrons, so that the information provides us a hint about electron distribution in the channel [11]. Although many experiments including our previous experiments carried out in the regime of $\nu_{\rm{qpc}}$ slightly less than $1$ and the typical RDNMR signals are conductance peak (resistance dip) \cite{Fauzi2019, Fauzi2018, Annisa2020}, the lineshapes changed depending on the experimental conditions and there is no systematic studies focusing on how the lineshapes change as a function of $\nu_{\rm{qpc}}$. In this paper, we focus to the RDNMR lineshapes of the QPC in the quantum Hall breakdown regime with $0 < \nu_{\rm{qpc}} <1$. We note that the breakdown refers to the QPC region with the filling factor less than 1.

\section{Experimental setup}

\begin{figure*}[t]
\begin{center}    
\centering
\includegraphics[width=
\linewidth]{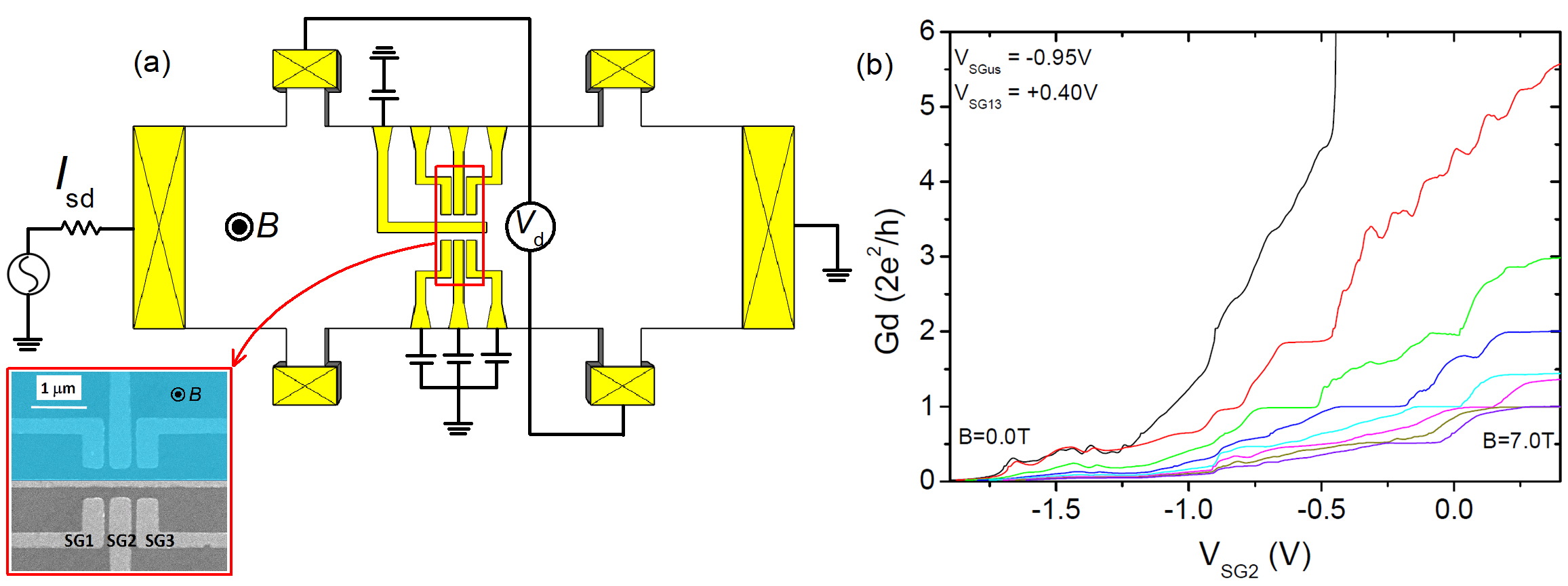}
\end{center}
\caption{(a) Device schematic and measurement setup. Inset shows a SEM image of multiple quantum point contacts. The negative bias is applied to the center gate and only the lower region is used in the experiments. SG2 is used to define the QPC. SG1 and SG3 are used to modify edge-channel landscape around the QPC. The device is cooled down to $300$ mK and subject to a magnetic field applied perpendicular to the device. (b) Conductance as a function of $V_{\rm{SG2}}$ measured at different magnetic fields from 0 to $7$ T with an interval of $1$ T, demonstrating a successful gate operation to control a number of edge channels transmitted through a single QPC.}
\label{Fig03} 
\end{figure*}

The experiment is carried out on a $20$-nm wide GaAs quantum well structure with the 2DEG located $140$ nm below the surface. We define a quantum point contact using multiple gate structures as shown schematically in Fig. \ref{Fig03}(a), enabling us to control the potential shape in the QPC saddle. The 2DEG electron density is field-induced by applying a positive voltage bias to backgate. The electron mobility is $\mu =2.8 \times 10^{5}$ cm$^2$/Vs at the electron density of $1.8 \times 10^{11}$ cm$^{-2}$. We deplete the upper half-section (blue-shaded) and use only the lower-half to define the QPC as shown in the inset of Fig. \ref{Fig03}(a). Throughout the measurement, we set the bulk filling factor to $\nu_{\rm{b}} = 2$ so that the spin-up and -down edge channels are available for transport. This is accomplished by threading the device with a $7$ Tesla magnetic field and tuning the electron density to $3.3 \times 10^{11}$ cm$^{-2}$, which corresponds to a backgate bias voltage of $5$ V. Applying a relatively high magnetic field of $7$ T ensures a large electronic Zeeman splitting and well-defined spin-resolved edge channel in our sample. We measure the diagonal voltage ($V_d$) to probe the number of edge channels transmitted through the constriction with a standard lock-in technique. The device is placed inside a single shot dry refrigerator and cooled down to $300$ mK.

Fig. \ref{Fig03}(b) displays basic transport characteristics of our device. We record the diagonal conductance ($G_{\rm{d}} = dI_{\rm{sd}}/dV_{\rm{d}}$) as a function of bias voltage applied to $\rm{SG2}$, measured from 0 to $7$ T perpendicular field with an interval of $1$ T. The bias voltage applied to $\rm{SG1}$ and $\rm{SG3}$ are held constant to $+0.4$ V and transparent to electrons. Due to a relatively low mobility GaAs wafer we use, we do not observe a sequence of quantized conductance at zero-field. However, a number of quantized conductance plateau can be observed when we turn the field on due to a large cyclotron energy gap. The number of modes available for transport below the Fermi level decreases as we increase the field as expected due to magneto-electric subbands depopulation \cite{Wees1988}. It reaches bulk filling factor $\nu_{\rm{b}} = 2$ at $7$ T, where only two spin edge channels available for transport.

The conductance decreases with applying more negative bias voltage to SG2, indicating a successful control of edge channel transmitted through the QPC. We have checked the 1D transport characteristic for the other two adjacent metal gates individually and observed similar magneto-conductance characteristics at a non-zero field displayed in Fig. \ref{Fig03}(b).

\section{Experimental results}

\subsection{Strong tunneling regime}

\begin{figure*}[t]
\begin{center}    
\centering
\includegraphics[width=
\linewidth]{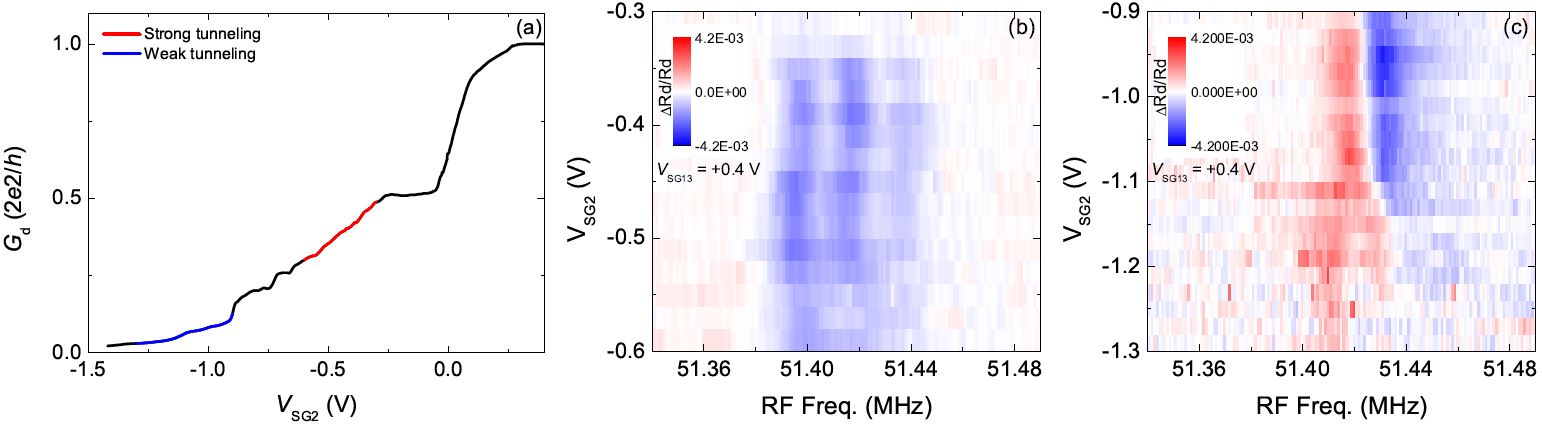}
\end{center}
\caption{(a) Magneto-conductance traces as a function of $V_{\rm{SG2}}$ measured at $7$ T and $300$ mK. SG1 and SG3 are set to $0.4$ V. The red and blue lines along the conductance traces indicate where continuous-wave RDNMR spectra are acquired. (b) $^{75}$As RDNMR spectra collected in a strong tunneling regime marked by the red trace in panel (a). (c) The spectra collected in a low tunneling regime marked by the blue conductance trace in panel (a). All spectra are swept with decreasing frequency with scan speed of $200$ Hz/s.}
\label{Fig04} 
\end{figure*}

We start off by showing $^{75}$As RDNMR spectra in the so-called strong tunneling regime along the red line indicated in Fig. \ref{Fig04}(a), where most studies are carried out. Here the constriction is defined by SG2 with both SG1 and SG3 are set to $+0.4$ V. We observe an usual resistance dip as expected when we set the local filling factor $\nu_{\rm{QPC}} < 1$. The resonance is split further into three resonance lines separated by about $20$ kHz due to quadrupole interaction between $^{75}$As nuclei and electric field gradient (EFG) \cite{Shulman1957, Sundfors1969}. Strain developed at the interface between the metal gates and GaAs semiconductor propagates down into the channel and becomes the major source of EFG \cite{Fauzi2019}. In a multiple gate architecture as is the case here, strain is expected to modulate spatially within the constriction. The modulation is reflected in the RDNMR spectrum we collect as displayed in Fig. \ref{Fig04}(b). A clear three-fold splitting becomes obscure and gets overlapped when the spectra are collected for $V_{\rm{SG2}} < -0.5$ V. This is because as we apply more a negative bias voltage to SG2, the channel is pushed toward the less strained region under the center gate. It clearly demonstrates that the forward spin-flip scattering events and hence dynamic nuclear polarization follows the channel movement. 

A closer look at the collected spectra shown in Fig. \ref{Fig04}(b) reveals another interesting feature. The center resonance line does not monotonically shift to lower frequency side with increasing a negative bias voltage to SG2, but rather it shows a snake-like pattern. In principle, when the nuclear spins are in contact with the electrons in the channel, the resonance line will be shifted to lower frequency due to Knight shift \cite{Chida2012, Fauzi2017, Hashisaka2020}. The collected Knight shift would follow the electron density distribution in the channel. Since the electron density in the QPC is linearly proportional to the applied gate voltage, we expect its Knight shift to follow the linear dependence with the applied gate voltage too, if the electron spins are fully polarized as expected from the simple theory of the integer quantum Hall regime. The snake-like feature we have observed indicates that the electron density within the QPC is spatially modulated although the SG1 and SG3 are set to be transparent. Furthermore, it was reported that the polarization of electron spin was complicated in the lowest Landau-level \cite{Tiemann2012}. We confirmed that such situation was kept at higher temperature around 1 K where fractional quantum feature disappeared \cite{Higashida}. Although we have not checked yet, similar situation might occur in the sample of a slightly lower mobility used here. The spatial modulation of electron distribution together with such unusual behavior of the electron spin polarization complicate the observed Knight shift. The detailed discussions need further studies and beyond the scope of this paper.

\begin{figure}[t]
\begin{center}    
\centering
\includegraphics[width=
\linewidth]{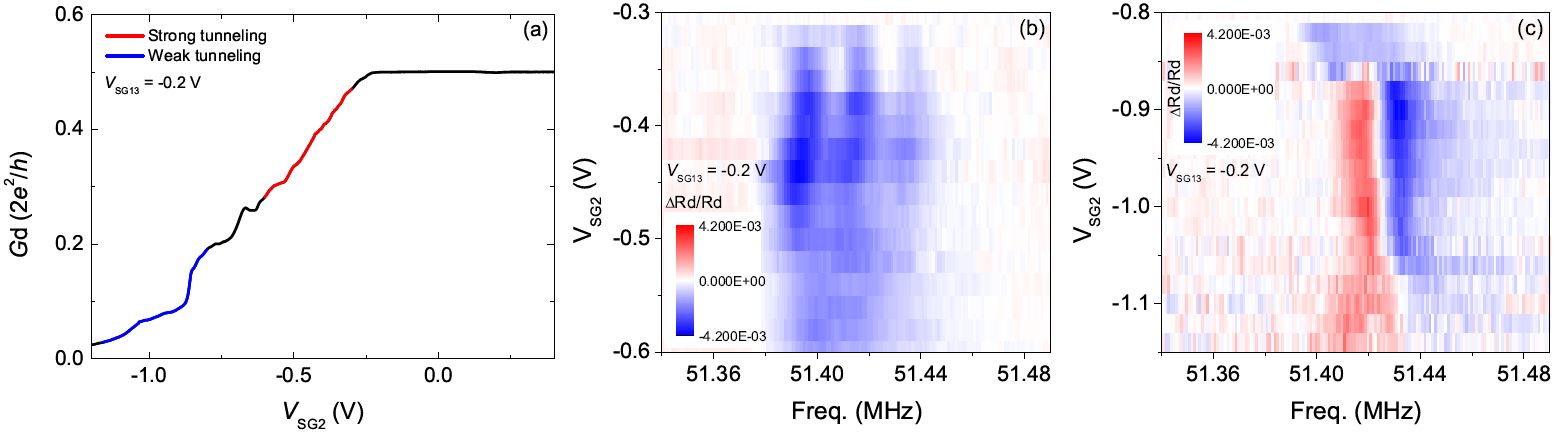}
\end{center}
\caption{(a) Magneto-conductance traces as a function of $V_{\rm{SG2}}$ measured at $7$ T and $300$ mK. SG1 and SG3 are set to $-0.2$ V. The red and blue lines along the conductance traces indicate where continuous-wave RDNMR spectra are acquired. (b) $^{75}$As RDNMR spectra collected in a strong tunneling regime marked by the red trace in panel (a). (c) The spectra collected in a low tunneling regime marked by the blue conductance traces in panel (a). All spectra are swept with decreasing frequency with scan speed of $200$ Hz/s.}
\label{Fig05}
\end{figure}

We now examine the impact of modifying the potential profile around SG2 by applying a negative bias voltage of $-0.2$ V to SG1 and SG3. We expect that the potential is strongly modulated around the QPC center. Since the electron density under SG1 and SG3 gates are now partially depleted, the down-spin channel gets reflected already from the beginning at $V_{\rm{SG2}} = +0.4$ V. This is apparent when we look at the conductance profile displayed in Fig. \ref{Fig05}(a), showing $G_d = 0.5 \times 2e^2/h$ at $V_{\rm{SG2}} = +0.4$ V. Interestingly, the spectra displayed in Fig. \ref{Fig05}(b) also exhibits the snake-like feature when the spectra are collected closer to the plateau at $-0.6 < V_{\rm{SG2}} < -0.3$ V. The similarity further confirms our idea that the feature is due to the spatial modulation of electron density in the QPC.

\subsection{Low tunneling regime: emergence of dispersive lineshapes}

Now let us move onto the RDNMR spectra collected in the low tunneling regime as highlighted by the blue line on the conductance traces displayed in Fig. \ref{Fig04}(a). Instead of the usual resistance dip, a dispersive lineshape emerges around $-0.9 < V_{\rm{SG2}} < -1.1$ V. A dip followed by a peak resonance line with decreasing the frequency with a dip-to-peak separation around $15$ kHz. Contrary to the previous case, the quadrupole splitting is clearly absent in the region where the dispersive lineshape is observed. It implies that the nuclear spins are polarized in a region with negligible strain field. While the resistance dip for $\nu_{\rm{QPC}} < 1$ is due to a positive nuclear spin polarization as we have discussed, the resistance peak arises due to a negative nuclear spin polarization from backward spin flip scattering events. Those two nuclear spin polarization domains are in contact with different electron spin polarization. The value of $15$ kHz that separates the two resonance lines reflects the Knight shift difference experienced by the two sets of nuclear spin polarization. Moving very close to the pinch off point, only the resistance peak arising from the negative nuclear spin polarization remains visible in the collected spectra. The resistance peak branches off into two resonance lines. This feature is not quadrupole interaction in origin but might reflect the Knight shift distribution in and/or near the QPC channel.

The same dispersive lineshape characteristic in the spectra is observed when we apply a negative bias voltage of $-0.2$ V to SG1 and SG3 as dispplayed in Fig. \ref{Fig05}(c) for $-1.1 < V_{\rm{SG2}} < -0.85$ V. The observed spectra resemble the one shown in Fig. \ref{Fig04}(c), however the intensity is doubly enhanced and the Knight shift is slightly bigger. The lineshape still emerges even though the QPC center and the $\nu = 2$ edge channel are widely separated in space.







\section{Conclusion}

We have observed RDNMR spectra collected in the strong and low tunneling regime in multiple quantum point contact devices. Depending where we are at, we find different sets of spectrum due to the manifestation of edge-tunneling between spin-up and -down electron in the quantum point contact. In the strong tunneling regime, the forward spin-flip scattering dominates the tunneling events leading to the build up of positive nuclear spin polarization. In the low tunneling regime, in addition to the forward type scattering, the backward type scattering can take place. It leads to the build up of two different sets of nuclear spin polarization and manifested in the dispersive lineshapes. The snake-like pattern in the RDNMR spectrum points to the presence of spatial modulation of electron density in the multiple quantum point contacts. 

\medskip

\textbf{Acknowledgements} 
We would like to thank K. Nagase and M. Takahashi for their help in fabricating the devices and measurements. K.H. and Y.H. thank financial support of Japan Society for the Promotion of Science KAKENHI 15H05867, 15K217270, 17H02728, 18H01811, and 26390006. M. H. F. and Y. H. also thank the support from Graduate Program in Spintronics, Tohoku University.
\medskip

\begin{thebibliography}{99}
\bibliographystyle{MSP}

\bibitem{Halperin1982} B.I. Halperin, \textit{Phys. Rev. B} \textbf{1982}, \textit{25}, 2185.

\bibitem{Haug1993} R J Haug, \textit{Semicond. Sci. and Technol.} \textbf{1993}, \textit{8}, 131.

\bibitem{Duprez2019} H. Duprez, E. Sivre, A. Anthore, A. Aassime, A. Cavanna, A. Ouerghi, U. Gennser, and F. Pierre, \textit{Phys. Rev. X} \textbf{2019}, \textit{9}, 021030.

\bibitem{Kane1992} B. E. Kane, L. N. Pfeiffer, and K. W. West, \textit{Phys. Rev. B} \textbf{1992}, \textit{46}, 7264(R).

\bibitem{Wald1994} K. R. Wald, L. P. Kouwenhoven, P. L. McEuen, N. C. van der Vaart, and C. T. Foxon, \textit{Phys. Rev. Lett}. \textbf{1994}, \textit{73}, 1011.

\bibitem{Dixon1997} D. C. Dixon, K. R. Wald, P. L. McEuen, and M. R. Melloch, \textit{Phys. Rev. B} \textbf{1997}, \textit{56}, 4743.

\bibitem{Hirayama2009} Y. Hirayama, G. Yusa, K. Hashimoto, N. Kumada, T. Ota, and K. Muraki, \textit{Semicond. Sci. Technol.} \textbf{2009}, \textit{24}, 023001.


\bibitem{Fauzi2019} M. H. Fauzi, M. F. Sahdan, M. Takahashi, A. Basak, K. Sato, K. Nagase, B. Muralidharan, and Y. Hirayama, \textit{Phys. Rev. B} \textbf{2019}, \textit{100}, 241301(R).

\bibitem{Fauzi2017} M. H. Fauzi, A. Singha, M. F. Sahdan, M. Takahashi, K. Sato, K. Nagase, B. Muralidharan, and Y. Hirayama, \textit{Phys. Rev. B} \textbf{2017}, \textit{95}, 241404(R).

\bibitem{Fauzi2018} M. H. Fauzi, A. Noorhidayati, M. F. Sahdan, K. Sato, K. Nagase, and Y. Hirayama, \textit{Phys. Rev. B} \textbf{2018}, \textit{97}, 201412(R).

\bibitem{Annisa2020} Annisa Noorhidayati, Mohammad Hamzah Fauzi, Muhammad Fauzi Sahdan, Shunta Maeda, Ken Sato, Katsumi Nagase, and Yoshiro Hirayama, \textit{Phys. Rev. B} \textbf{2020}, \textit{101}, 035425.

\bibitem{Desrat2002} W. Desrat, D. K. Maude, M. Potemski, J. C. Portal, Z. R. Wasilewski, and G. Hill, \textit{Phys. Rev. Lett.} \textbf{2002}, \textit{88}, 256807.

\bibitem{Tracy2006} L. A. Tracy, J. P. Eisenstein, L. N. Pfeiffer, and K. W. West, \textit{Phys. Rev. B} \textbf{2006}, \textit{73}, 121306.

\bibitem{Kodera2006} K. Kodera, H. Takado, A. Endo, S. Katsumoto, and Y. Iye, \textit{Physica Status Solidi (c)} \textbf{2006}, \textit{3}, 4380.

\bibitem{Dean2009} C. R. Dean, B. A. Piot, G. Gervais, L. N. Pfeiffer, and K. W. West, \textit{Phys. Rev. B} \textbf{2009}, \textit{80}, 153301.

\bibitem{Bowers2010} C. R. Bowers, G. M. Gusev, J. Jaroszynski, J. L. Reno, and J. A. Simmons, \textit{Phys. Rev. B} \textbf{2010}, \textit{81}, 073301.

\bibitem{Desrat2013} W. Desrat, B. A. Piot, S. Kr¨amer, D. K. Maude, Z. R. Wasilewski, M. Henini, and R. Airey, \textit{Phys. Rev. B} \textbf{2013}, \textit{88}, 241306.

\bibitem{Desrat2015} W. Desrat, B. A. Piot, D. K. Maude, Z. R. Wasilewski, M. Henini, and R. Airey, \textit{Journal of Physics: Condensed Matter} \textbf{2015}, \textit{27}, 275801.

\bibitem{Piot2016} B. A. Piot, W. Desrat, D. K. Maude, D. Kazazis, A. Cavanna, and U. Gennser, \textit{Phys. Rev. Lett.} \textbf{2016}, \textit{116}, 106801.

\bibitem{Keane2011} Z. K. Keane, M. C. Godfrey, J. C. H. Chen, S. Fricke, O. Klochan, A. M. Burke, A. P. Micolich, H. E. Beere,
D. A. Ritchie, K. V. Trunov, D. Reuter, A. D. Wieck,
and A. R. Hamilton, \textit{Nano Letters} \textbf{2011}, \textit{11}, 3147.

\bibitem{Chida2012} K. Chida, M. Hashisaka, Y. Yamauchi, S. Nakamura, T. Arakawa, T. Machida, K. Kobayashi, and T. Ono, \textit{Phys. Rev. B} \textbf{2012}, \textit{85}, 041309.

\bibitem{Okazaki2018} Y. Okazaki, I. Mahboob, K. Onomitsu, S. Sasaki, S. Nakamura, N. Kaneko, and H. Yamaguchi, \textit{Nat. Commun.} \textbf{2018}, \textit{9}, 2993.

\bibitem{Hashisaka2020} Masayuki Hashisaka, Koji Muraki, and Toshimasa Fujisawa, \textit{Phys. Rev. B} \textbf{2020}, \textit{101}, 041303(R).

\bibitem{Wees1988} B. J. van Wees, L. P. Kouwenhoven, H. van Houten, C. W. J. Beenakker, J. E. Mooij, C. T. Foxon, and J. J. Harris, \textit{Phys. Rev. B} \textbf{1988} \textit{38}, 3625(R).

\bibitem{Wees1991} B. J. van Wees, L. P. Kouwenhoven, E. M. M. Willems, C. J. P. M. Harmans, J. E. Mooij, H. van Houten, C. W. J. Beenakker, J. G. Williamson, and C. T. Foxon, \textit{Phys. Rev. B} \textbf{1991}, \textit{43}, 12431.

\bibitem{Shulman1957} R. G. Shulman, B. J. Wyluda, and P. W. Anderson, \textit{Phys. Rev}. \textbf{1957}, \textit{107}, 953.

\bibitem{Sundfors1969} R. K. Sundfors, \textit{Phys. Rev}. \textbf{1969}, \textit{177}, 1221.




\bibitem{Singha2017} Aniket Singha, M. H. Fauzi, Yoshiro Hirayama, Bhaskaran Muralidharan. \textit{Phys. Rev. B} \textbf{2017}, \textit{95}, 115416,.

\bibitem{Tiemann2012} L. Tiemann, G. Gamez, N. Kumada, and K. Muraki, \textit{Science} \textbf{2012}, \textit{335}, 828.

\bibitem{Higashida} R. Higashida, S. Hasegawa, K. Hashimoto, T. Kobyashi, and Y. Hirayama, unpublished.

\bibliography{MSP-template}
\end{thebibliography}
\end{document}